\documentclass[preprint,superscriptaddress]{revtex4}
\usepackage{graphicx}
\usepackage{dcolumn}
\usepackage{bm}
\usepackage{amsmath,amsthm,amssymb}
\begin{document}
%


\title{How to bypass Birkhoff through extra dimensions\\
{\small A simple framework for investigating the gravitational
collapse in vacuum}}

\author{Piotr Bizo\'n}
 \affiliation{M. Smoluchowski Institute of Physics, Jagiellonian University,
Reymonta 4, 30-059 Krak\'ow, Poland}\affiliation{Max-Planck-Institut
f\"ur Gravitationsphysik, Albert-Einstein-Institut, Am M\"uhlenberg
1, D-14476 Golm, Germany \vspace{0.4cm}}
\author{Bernd G. Schmidt}
    \affiliation{Max-Planck-Institut f\"ur Gravitationsphysik,
Albert-Einstein-Institut, Am M\"uhlenberg 1, D-14476 Golm, Germany
\vspace{0.4cm}}
\date{\today \\ \vspace{2cm}}
\begin{abstract}
It is fair to say that our current mathematical understanding of the
dynamics of gravitational collapse to a black hole is limited to the
spherically symmetric situation and, in fact, even in this case much
remains to be learned. The reason is that Einstein's equations
become tractable only if they are reduced to a $1+1$ dimensional
system of partial differential equations. Due to this technical
obstacle, very little is known about the collapse of pure
gravitational waves because by Birkhoff's theorem there is no
spherical collapse in vacuum.

 In this essay we describe a new cohomogeneity-two
symmetry reduction of the vacuum Einstein equations in five and
higher odd dimensions which evades Birkhoff's theorem and admits
time dependent asymptotically flat solutions. We argue that this
model
  provides an attractive $1+1$ dimensional geometric setting for investigating the
  dynamics of gravitational collapse in vacuum.
\end{abstract}

\maketitle
The understanding of the nature of singularities that arise from
gravitational collapse is one of the most challenging problems in
classical general relativity. The key open issue is to resolve the
so called  weak cosmic censorship hypothesis which says that in
reasonable models of gravitational collapse naked singularities do
not generically develop from regular initial data \cite{p}. If this
hypothesis holds, the ignorance of laws of physics near the
singularity does not prevent a distant observer from applying
ordinary physics outside black holes.

 An attempt to prove the cosmic censorship hypothesis in full
generality is beyond the scope of existing mathematical techniques
so the attention of researchers has been focused on more tractable
special cases, in particular spherically symmetric spacetimes.
Although the assumption of spherical symmetry greatly simplifies the
problem, it has a drawback of becoming trivial in vacuum since, by
Birkhoff's theorem, the spherical gravitational field has no
dynamical degrees of freedom. Thus, in order to generate dynamics it
is necessary to choose a matter model. A popular choice, which has
led to important insights, is a real massless scalar field. For this
matter model Christodoulou showed that for small initial data the
fields disperse to infinity \cite{ch1}, while for large initial data
black holes are formed \cite{ch2}. The transition between these two
scenarios was explored numerically by Choptuik \cite{ch} leading to
the discovery of critical phenomena at the threshold of black hole
formation. Later studies showed that the character of gravitational
collapse and critical phenomena depends both qualitatively and
quantitatively on a matter model \cite{g} -- for this reason it is
difficult to determine which features of spherical collapse are
model-dependent and which ones hold in general.

In this essay we present a new framework which -- at the expense of
going to five (or
 higher odd)
 dimensions --
evades Birkhoff's theorem and admits radially symmetric
gravitational waves . For the sake of clarity, we shall illustrate
the main idea in five dimensions and only at the end we shall
briefly discuss higher dimensional generalizations.

We begin by recalling some elementary aspects of the geometry of
spheres, in particular the three-sphere $S^3$. Usually, when  one
thinks of a sphere, one imagines a round sphere, however there are
other natural geometries with which a manifold $S^n$ can be endowed.
To see this, note that an $n$-sphere is a homogeneous manifold, that
is a manifold which looks the same at every point. More formally, a
manifold is said to be homogeneous if there is a Lie group which
acts transitively on it. Any $n$-sphere is homogeneous because its
group of rotations $SO(n+1)$ acts transitively on it. On $S^2$ there
is no other transitive action than $SO(3)$, however some higher
dimensional spheres  admit transitive actions of a proper subgroup
of the rotation group. In particular, it is well known that the
group $SU(2)$ acts transitively on $S^3$. This is most easily seen
by considering a sequence of rotations acting on the standard
three-sphere in four-dimensional Euclidean space.  A simultaneous
rotation in the $xy$ and $zw$ planes keeps no points on $S^3$ fixed.
It is geometrically evident that this rotation, together with the
simultaneous rotations in the $xz,yw$ and $xw,yz$ planes,  forms a
simply transitive group $G$ acting on $S^3$. By computing the Lie
algebra of $G$  it is easy to verify that $G$ is isomorphic to
$SU(2)$. Given a bare homogeneous manifold it is natural to endow it
with a metric which is invariant under the transitive symmetry group
-- such a metric is called homogeneous. It follows from the above
that on $S^3$ there are two possible homogeneous metrics: the
standard maximally symmetric $SO(4)$-invariant metric and the less
symmetric $SU(2)$-invariant metric. The latter one has the form
\begin{equation}\label{s3}
ds^2 = L_1^2 \sigma_1^2 + L_2^2 \sigma_2^2 + L_3^2 \sigma_3^2,
\end{equation}
where $\sigma_k$ are left-invariant one-forms on $SU(2)$.
 In terms of Euler
angles,
\begin{equation}\label{sigma}
    \sigma_1=\cos{\psi}\;d\theta+\sin{\psi}\sin{\theta}\;d\phi,\quad
    \sigma_2=-\sin{\psi}\;d\theta+\cos{\psi}\sin{\theta}\;d\phi,\quad
     \sigma_3=d\psi + \cos{\theta}\; d\phi.
\end{equation}
The constant coefficients $L_k$ have the interpretation of the
principal curvature radii of $S^3$.  If all three $L_k$ are equal,
the metric (\ref{s3}) reduces to the standard round metric,
otherwise the metric is anisotropic and the three-sphere is said to
be "squashed". The metric (\ref{s3}) might be familiar from the
mechanics of rigid bodies where it appears as the kinetic energy
metric  of an asymmetric top (with $L_k^2$ being the principal
moments of inertia). In relativity, the metric (\ref{s3}) is well
known from the studies of the Bianchi IX homogeneous cosmological
models but, strangely enough, a possibility of using it in the
context of gravitational collapse seems to have been overlooked.

Now, the key idea is to use (\ref{s3}) as the angular part of a
cohomogeneity-two metric in five spacetime dimensions. More
precisely, we consider the five dimensional spacetime which is
foliated by the squashed three-spheres endowed with the
$SU(2)$-invariant homogeneous metrics proportional to (\ref{s3}).
Using the coordinate freedom in the two-space orthogonal to the
$S^3$ group orbit of $SU(2)$, we choose the volume radial coordinate
$r=(vol(S^3)/2\pi^2)^{1/3}$, and write the spacetime metric in the
following form
\begin{equation}\label{metric}
    ds^2= - A e^{-2\delta} dt^2 + A^{-1} dr^2 + \frac{1}{4} r^2 \left[ e^{2B}
    \sigma_1^2+e^{2C} \sigma_2^2 +e^{-2(B+C)}\sigma_3^2\right],
\end{equation}
where $A$, $\delta$, $B$, and $C$ are functions of $t$ and $r$.
Substituting the ansatz (\ref{metric}) into the vacuum Einstein
equations in five dimensions we get  equations of motion for the
functions $A, \delta$, $B$, and $C$. These equations are rather
complicated, so here we make an additional simplifying assumption
that $B=C$ which means that the ansatz (\ref{metric}) has an extra
$U(1)$ symmetry.

 Using the mass function $m(t,r)$, defined by $A=1-m/r^2$,
we obtain the following equations
\begin{subequations}
\begin{eqnarray}\label{const}
 m' & = & 2r^3\left(e^{2\delta} A^{-1} {\dot B}^2 + A {B'}^2\right)+\frac{2}{3} r
 \left(3+ e^{-8B}-4 e^{-2B}\right),\\
\delta' &=& - 2 r \left(e^{2\delta} A^{-2}{\dot B}^2 +
    B'^2\right),
    \end{eqnarray}
\begin{equation}\label{wave}
 \left(e^{\delta} A^{-1} r^3 {\dot B}\right)^{\cdot} -
\left(e^{-\delta} A r^3 B'\right)' + \frac{4}{3} e^{-\delta} r
\left(e^{-2B}-e^{-8B}\right)=0,
\end{equation}
\end{subequations}
where primes and dots denote derivatives with respect to $r$ and
$t$, respectively. It is evident from these equations that the only
dynamical degree of freedom  is the field $B$
 which plays a role similar to
a matter field in the spherically symmetric Einstein-matter systems
in four dimensions. If $B=0$ we have spherical symmetry and, in
agreement with Birkhoff, the only solutions are Minkowski
($\delta=0$, $m=0$) and Schwarzschild ($\delta=0$, $m=const>0$).

The system (4) provides the simplest possible setting for
investigating
  the dynamics of gravitational collapse in vacuum.
%
   Numerical simulations
   indicate that the spherically symmetric solutions, Minkowski and Schwarzschild,
   play the role of attractors in the evolution
  of  generic regular initial data  (small and large ones,
  respectively) and the transition between
these two  outcomes of evolution exhibits the type II discretely
self-similar critical behavior \cite{bcs}. These results strongly
suggest that the above mentioned Christodoulou's proof of the weak
cosmic censorship hypothesis for the self-gravitating scalar field
can be repeated for our system. Actually, the first step in this
direction has already been done by Dafermos and Holzegel \cite{dh}
who proved nonlinear stability of the Schwarzschild solution for
perturbations respecting the ansatz (\ref{metric}).

There are many other natural problems that can be addressed within
the proposed framework - due to space limitations we mention only
some of them:
\begin{itemize}
\item \emph{Asymptotic stability of the Schwarzschild black hole:}
It is well known from  heuristic and numerical studies that
convergence to the Schwarzschild black hole proceeds first through
exponentially damped oscillations (quasinormal modes) and then
through a polynomial decay (tail), however a mathematical
description of this relaxation process remains elusive. Since
dissipation by dispersion is much more effective in higher
dimensions, the system (4) provides a promising
setup~for~studying~this~problem.
\item \emph{Critical collapse:} The most mysterious aspect  of Choptuik's discovery \cite{ch} is the
  discrete self-similarity of the critical solution. The fact that
  the critical solution in our model also has this strange symmetry lends support to the belief
  that it  is an intrinsically  gravitational phenomenon and the typical feature
  of critical collapse.  We believe that our model might be helpful in unraveling
   a mechanism which is responsible for the discrete self-similarity
of the critical solution.
  We remark in passing
  that the system (4) does not admit regular continuously self-similar solutions -- this indicates that
  the  naked singularities of this kind found by Christodoulou for the self-gravitating
  massless scalar field \cite{ch3} are, in a sense, matter-generated and as
  such
  much less interesting.

   Obviously, the relevance of critical phenomena
  depends on how generic they are. Our numerical results \cite{bcs}
  show that they are not restricted to spherically~symmetric
  Einstein-matter systems but occur also in the collapse of pure gravitational~waves~\footnote{The evidence for
critical behavior in the collapse of axisymmetric
  gravitational waves given by Abrahams and Evans in \cite{ae} is much less  compelling because of numerical difficulties.}.
  It is worth to point out that an extra $U(1)$ symmetry that we have assumed
  to simplify the ansatz (\ref{metric}) does not effect the
  critical collapse: the  evolution of critical initial data with two radiative
  degrees of freedom $B$ and $C$ tends to the same critical
  solution as in the simplified model, in other words the $U(1)$ symmetry
  is recovered dynamically \cite{bcs3}. The next important step in
  determining the genericity of the critical behavior would be
be to analyse  the  linear stability of the critical solution
against general perturbations.
\item \emph{Non-asymptotically flat solutions:} The ansatz
(\ref{metric}) is very rich and incorporates as special cases many
non-asymptotically flat  metrics~\footnote{We thank Gary Gibbons for
pointing this out.}. In particular, some gravitational instantons,
i.e. $4$-dimensional Riemannian Ricci-flat metrics (like for
instance the Taub-NUT metric),  can be viewed as static solutions in
our model.   In a sense, one can say that these solutions were
waiting for the equations -- having the system (4), it becomes
possible to ask questions  about the dynamical role of gravitational
instantons and their stability properties \cite{bcg}.

\item \emph{Higher-dimensional generalizations:} It should be clear from the  derivation of the ansatz
(\ref{metric}) that similar models can be formulated in higher
$D=n+2$ dimensions as long as the corresponding $n$-sphere  admits a
non-round homogeneous metric, i.e. there exists a proper subgroup of
the rotation group $SO(n+1)$ which acts transitively on $S^n$.
According to the classification given by Besse \cite{besse} such
transitive actions exist on all odd dimensional spheres. For
example, the group $SU(n+1)$ acts transitively on $S^{2n+1}$ and the
group $Sp(n+1)$ acts transitively on $S^{4n+3}$. It is natural to
ask whether the properties of gravitational collapse found in
\cite{bcs} are typical for this class of models or whether new
phenomena appear in higher dimensions. Having this motivation in
mind, together with our collaborators we have recently studied  the
nine dimensional cohomogeneity-two model based on the squashed
seven-sphere with $SO(8)$ isometry broken to $SO(5)\times SU(2)$
\cite{bcs2}. We found that the overall picture of gravitational
collapse and critical phenomena is qualitatively the same as in five
dimensions, except for one  surprising difference:  while in five
(and, of course, four) dimensions the mass function $m(t,r)$ is
monotone increasing with $r$ (as follows immediately from equation
(4a)),  in nine dimensions the mass function can decrease. This
means that the energy density of the gravitational field may be
locally negative and suggests a possibility of violating the weak
cosmic censorship. This intriguing problem is being investigated.
Preliminary studies show no evidence for naked singularities -- the
negative energy regions seem to be always eventually shielded by
horizons. If this happens in general, it would indicate that the
cosmic censor is more powerful than it has been thought.
\end{itemize}
To summarize,  we have proposed a new theoretical framework for
investigating gravitational collapse in a clean matter-free
environment. We think that this framework is rich enough to capture,
at least in part, the general features of vacuum Einstein's
equations. We would like to remark that although our own motivation
for going to higher dimensions  was mathematical in nature, nowadays
extra dimensions are  very common in physics, in particular the
string theory predicts that our world has  more than four
dimensions.
 In order to
understand the phenomenology of various string-related  models (such
as the brane-world scenario) and their experimental predictions,
like production of mini black holes in the next generation of
colliders, it is important to study solutions of
 Einstein's equations in higher  dimensions. From this
 perspective the ideas presented in this essay may turn out to be quite
 useful.

\end{document}